\documentclass[10pt,twocolumn]{article}
\pdfoutput=1
\usepackage{cite}
\usepackage{amsmath,bbm}
\usepackage{amssymb}
\usepackage{graphicx} \graphicspath{{images/}}
\usepackage[latin1]{inputenc}
\usepackage{mathrsfs}
\usepackage{mathtools}
\usepackage[dvipsnames]{xcolor}
\usepackage[footnotesize]{caption}
\usepackage{xcolor}
\usepackage{hyperref}
\usepackage[hmarginratio=1:1,top=32mm,columnsep=20pt]{geometry}
\usepackage{paralist}
\usepackage{abstract}
\usepackage{upgreek}

    \makeatletter
    \let\@fnsymbol\@arabic
    \makeatother

\newcommand{\bea}{\begin{eqnarray}}
\newcommand{\eea}{\end{eqnarray}}
\newcommand{\be}{\begin{equation}}
\newcommand{\ee}{\end{equation}}
\newcommand{\ba}{\begin{array}}
\newcommand{\ea}{\end{array}}

\def\gsim{\mathrel{\rlap{\lower4pt\hbox{\hskip1pt$\sim$}}
    \raise1pt\hbox{$>$}}}

\title{\vspace{-15mm}
	\fontsize{16pt}{10pt}\selectfont
	\textbf{Sterile neutrino searches via displaced vertices at LHCb}
	}	
\author{%
	\large
	\textsc{Stefan~Antusch$^{\star \dagger}$
	,  Eros~Cazzato$^\star$
	, Oliver~Fischer$^\star$
	}\\[10pt]
	\normalsize	$^\star$ Department of Physics, University of Basel, \\ 
	\normalsize 	Klingelbergstr.\ 82, CH-4056 Basel, Switzerland\\[5pt]
	\normalsize	$^\dagger$ Max-Planck-Institut f\"ur Physik (Werner-Heisenberg-Institut),\\
	\normalsize	F\"ohringer Ring 6, D-80805 M\"unchen, Germany
	\vspace{-5mm}
	}
\date{}

\begin{document}

\maketitle

\begin{abstract}
\noindent 
We explore the sensitivity of displaced vertex searches at LHCb for testing sterile neutrino extensions of the Standard Model towards explaining the observed neutrino masses. We derive estimates for the constraints on sterile neutrino parameters from a recently published displaced vertex search at LHCb based on run 1 data. They yield the currently most stringent limit on active-sterile neutrino mixing in the sterile neutrino mass range between 4.5 GeV and 10 GeV. Furthermore, we present forecasts for the sensitivities that could be obtained from the run 2 data and also for the high-luminosity phase of the LHC.
\\

\end{abstract}

\noindent
\paragraph{Introduction:}

It is known from neutrino oscillation experiments that at least two of the neutrino degrees of freedom of the Standard Model (SM) are massive. The masses of the light neutrinos may be generated by extending the SM with at least two so-called sterile (``right-handed'') neutrinos which can have a Majorana mass as well as Yukawa couplings to the three active neutrinos and to the Higgs doublet.
After the electroweak symmetry is broken the sterile and active neutrinos mix, resulting in light and heavy mass eigenstates which all participate in the weak interactions. 

When the sterile neutrinos are lighter than the $W$ boson and have small mixings, they can be long-lived such that their decay happens displaced from the primary interaction vertex. Such displaced vertex signatures, in a more general context, are searched for by the LHC collaborations, see for instance \cite{Aad:2014yea,CMS:2014wda,ATLAS:2017bvh}. 
Recently, the LHCb collaboration published their results from a search for massive long-lived particles decaying into $\mu j j$ \cite{Aaij:2016xmb}, which, as we will discuss below, are very powerful for constraining sterile neutrino parameters. 

In \cite{Aaij:2016xmb}, the datasets from run~1 for the center-of-mass energies of 7 and 8 TeV with luminosities of 1 and 2 fb$^{-1}$, respectively, are analysed. 
It was found that displaced vertices in the cylindrical LHCb sub-detector, the so-called VErtex LOcator (VELO), with a transverse displacement of up to 2 cm, are in agreement with the SM expectation, and no event was found with a transverse displacement larger than $5$ mm when applying the lepton isolation filter. 
For transverse displacements larger than 2 cm no events were recorded at all. 

In the following we use the results from \cite{Aaij:2016xmb} to derive estimates for the present constraints on sterile neutrino parameters.
Furthermore, we present forecasts for the possible sensitivities from a similar analysis of the LHCb run 2 data and also for the high-luminosity phase of the LHC.

\paragraph{Benchmark model:} 
As benchmark model we consider the ``symmetry protected seesaw scenario'' (SPSS), see for instance in ref.~\cite{Antusch:2015mia} and references therein, which features a pair of sterile neutrinos that are subject to a 
``lepton number''-like symmetry (and possible additional sterile neutrinos which are decoupled from collider phenomenology).  

The resulting mass eigenstates after electroweak symmetry breaking are the three light neutrinos $\nu_i$ $(i=1,2,3)$, which are massless, and two heavy neutrinos $N_j$ $(j=1,2)$ with approximately degenerate mass eigenvalues $M$.
The mass eigenstates are admixtures of the active (SM) and sterile neutrinos, and the neutrino mixing can be quantified by the mixing angles and their magnitude:
\begin{equation}
\theta_\alpha = \frac{y_{\nu_\alpha}^{*}}{\sqrt{2}}\frac{v_\mathrm{EW}}{M}\,, \qquad |\theta|^2 := \sum_{\alpha} |\theta_\alpha|^2\,,
\label{def:thetaa}
\end{equation}
with  $v_\mathrm{EW} = 246.22$ GeV, and where the $y_{\nu_\alpha}$ are the neutrino Yukawa couplings.
The observed light neutrino masses can be generated by a slight breaking of the protective symmetry, allowing the $y_{\nu_\alpha}$ to be in principle large for any given heavy neutrino mass $M$.
In this model, the mixing angles $\theta_{\alpha}$ and the heavy neutrino mass scale $M$ are essentially free parameters, with $M$ potentially around the EW scale.

\begin{figure}[b]
\begin{center}
\includegraphics[width=0.4\textwidth]{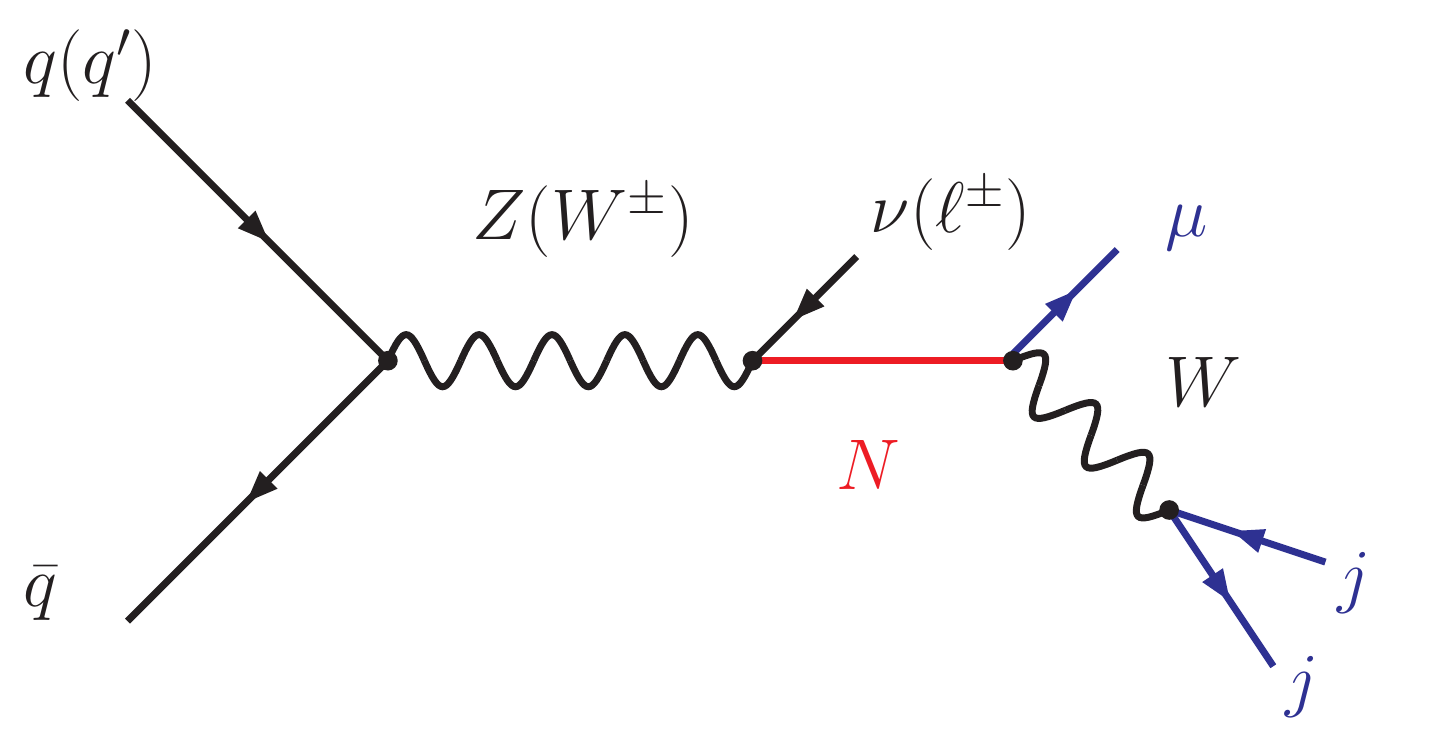}
\end{center}
\caption{Feynman diagram for the dominant signal processes that give rise to a displaced vertex with emergent $\mu j j$ final state from the decay of the heavy neutrino $N$.}
\label{fig:Nproduction}
\end{figure}

\paragraph{Displaced sterile neutrino decays at LHCb: }
As in LHCb's analysis \cite{Aaij:2016xmb}, we are considering the $\mu j j$ final state that emerges from a displaced secondary vertex. The most relevant diagrams which lead to this final state in the presence of sterile neutrinos are depicted in fig.\ \ref{fig:Nproduction}, with the dominant production channel being Drell-Yan.
For masses $M$ below $m_W$ and small $|\theta|^2$ the lifetime of the heavy neutrino can be long enough such that its decays give rise to a secondary vertex with a macroscopic displacement from the interaction point, where it has been produced. The non-observation of displaced vertices at LHCb can be used to constrain the sterile neutrino parameters, i.e.\ $M$ and $|\theta|^2$.

Based on geometric considerations, the expected number of signal events from the displaced decays of heavy neutrinos, $N_{\rm dv}$, which is subject to constraints from the LHCb analysis, is given by:
\begin{align}
&N_{\rm dv}(\sqrt{s},{\cal L},M,|\theta|^2) 
= 
\,\sum_{\text{x}{=}\nu,\ell^\pm} \overbrace{
\sigma_{\text{x} N}(\sqrt{s},M,|\theta|^2)\,{\rm Br}_{\mu jj}\, {\cal L}}^{N_{\text{x}N}}\times
\nonumber
\\
&\int D_{\text{x}N}(\vartheta,\gamma)\,P_{\rm dv}(x_{\rm min}(\vartheta),x_{\rm max}(\vartheta),\Delta x_{\rm lab}(\uptau,\gamma))\, d\vartheta d\gamma\,.
\label{eq:masterequation}
\end{align}
In this equation, $\sigma_{\text{x} N}$ labels the cross section for the different production processes with $\text{x}=\{\nu,\ell^\pm\}$ depicted in fig.~\ref{fig:Nproduction}, Br$_{\mu jj}$ is the branching ratio of a heavy neutrino into a muon plus jets, ${\cal L}$ denotes the integrated luminosity, which yields the overall number $N_{\text{x}N}$ of events with a displaced secondary vertex from heavy neutrino decays.
The integral gives the fraction of the heavy neutrino events that fall into the geometric acceptance of the LHCb detector.
Therein $D_{\text{x}N}(\vartheta,\gamma)$ is the probability distribution of producing a heavy neutrino with an angle $\vartheta$ between its momentum and the beam axis and Lorentz boost $\gamma$ for the given production process, and $P_{\rm dv}$ is the probability distribution of a decay to occur within a minimal and a maximal distance from the interaction point $x_{\rm min}$ and $x_{\rm max}$, respectively. Eq.~\eqref{eq:masterequation} goes beyond the treatment in ref.~\cite{Antusch:2016vyf} by including the detector geometry and the process specific kinematics. 
\begin{figure}[!b]
\begin{center}
\includegraphics[width=0.25\textwidth]{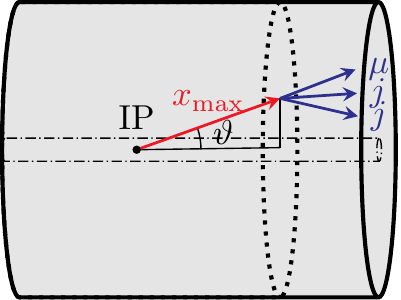}
\end{center}
\caption{The direction dependent displacement is shown for the example of sterile neutrinos decaying inside the sub-detector of LHCb, the vertex locator VELO.}
\label{fig:velo}
\end{figure}

The direction dependent maximal and minimal displacements are set by the geometry of the LHCb detector, in particular by the maximal longitudinal and transverse displacement, $z_{\rm max}$ and $r_{\rm max}$, respectively, whereas the minimal transverse displacement $r_{\rm{min}}$ is set by LHCb's analysis of the background. The resulting maximal displacement and minimal displacement are given by:

\begin{equation*}
x_{\rm max}(\vartheta) = 
\begin{cases*}
\frac{z_{\rm{max}}}{\cos(\vartheta)} & if $0\leq\vartheta\leq \arctan(r_{\rm max} / z_{\rm{max}} )$ \\
\frac{r_{\rm max}}{\sin(\vartheta)}  & if $\arctan(r_{\rm max}/z_{\rm{max}} ) \leq \vartheta < \frac{\pi}{2}$
\end{cases*},
\end{equation*}

\begin{equation}
x_{\rm min}(\vartheta) = 
\begin{cases*}
\frac{r_{\rm{min}}}{\sin(\vartheta)} & if $\arctan(r_{{\rm{min}}}/z_{\rm{max}})\leq \vartheta<\frac{\pi}{2}$\\
\mbox{n.a.}  & otherwise
\end{cases*}.
\label{eq:rmax}
\end{equation}
The used values for $z_{\rm{max}},$ $r_{\rm max}$ and $r_{\rm{min}}$ will be discussed in the next section, we show in fig.~\ref{fig:velo} an example for displaced decays inside the VELO.

\begin{figure}
\centering
\includegraphics[width=0.4\textwidth]{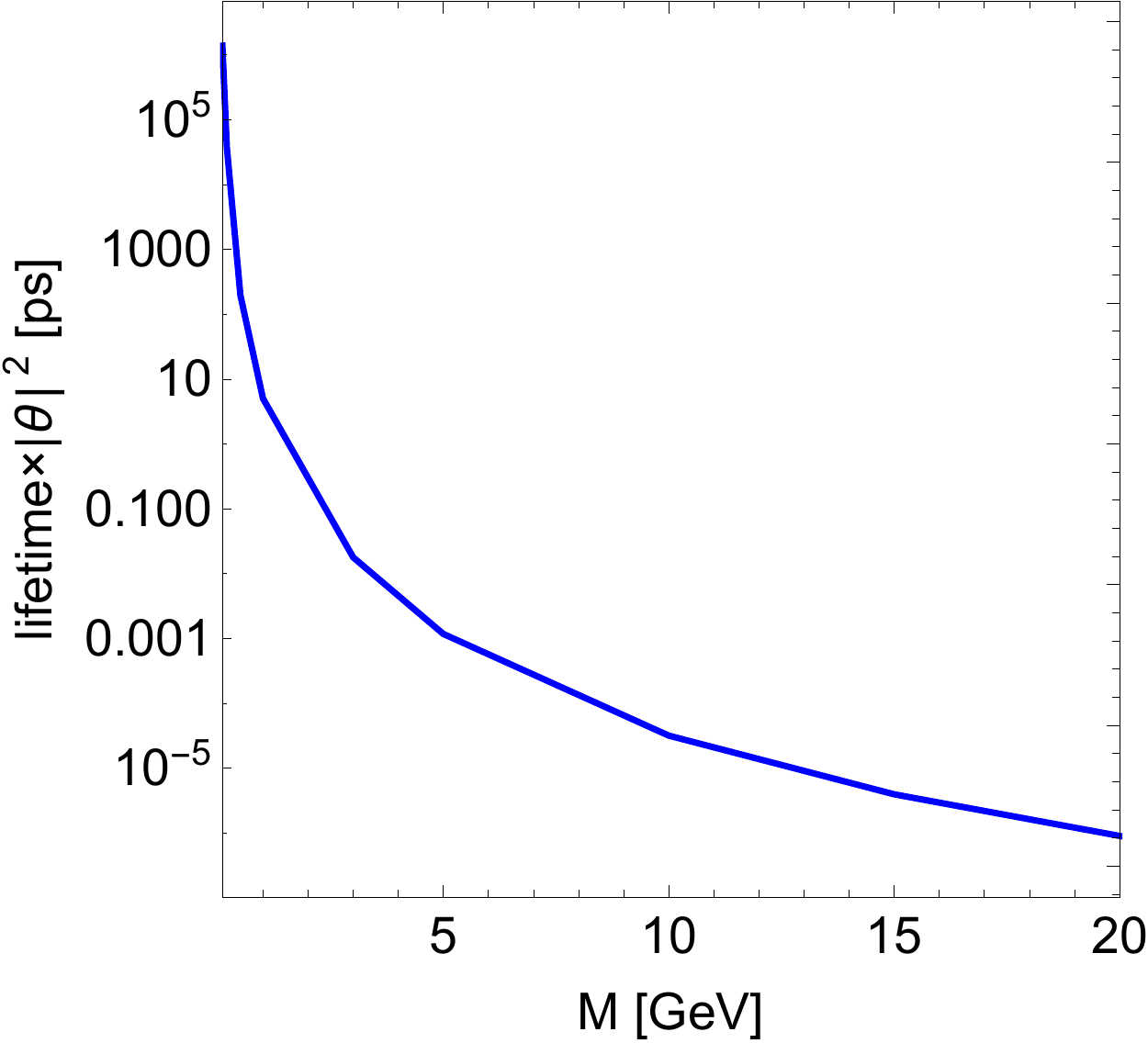}
\caption{Lifetime (in picoseconds) of a heavy neutrino multiplied with the active-sterile mixing parameter $|\theta|^2$ as a function of its mass.}
\label{fig:lifetime}
\end{figure}

The probability $P_{\rm dv}$ for a heavy neutrino to decay within these distances is dependent on the lifetime $\uptau$ of the heavy neutrino, here proper lifetime, and the corresponding Lorentz boost. 
The proper lifetime, shown in fig.\ \ref{fig:lifetime}, is obtained from the heavy neutrino width $\Gamma_N$, which is dependent on the active-sterile mixing $|\theta|^2$ and the mass $M$, by $\uptau=1/\Gamma_N$.
The probability of particle decays is exponentially distributed and for a displacement $\Delta x_{\rm lab}$ from the primary vertex with $x_{\rm min} \leq \Delta x_{\rm lab} \leq x_{\rm max}$ this probability is given in the laboratory frame by
\begin{equation}
P_{\rm dv}= {\rm Exp}\left(\frac{-x_{\rm min}}{\Delta x_{\rm lab}}\right) -  {\rm Exp}\left(\frac{-x_{\rm max}}{\Delta x_{\rm lab}}\right)\,.
\label{eq:probability}
\end{equation}
In the laboratory frame the displacement is
\begin{equation}
\Delta x_{\rm lab} = \uptau_{\rm lab} |\vec v| = \sqrt{\gamma^2-1}\, \uptau c\,,
\end{equation}
with $c$ the speed of light and $\gamma = \sqrt{1+ |\vec p|^2/M^2}$ being the Lorentz factor with the three-momentum of the heavy neutrino in the laboratory frame $\vec p$.

Now we have assembled all the ingredients to constrain the sterile neutrino parameters: 
The rare process of a displaced vertex decay is taken to be a Poisson process, thus the number of events are Poisson distributed with expected value $N_{\rm dv}$ as defined in eq.\ \eqref{eq:masterequation}. 
The non-observation of displaced vertex events puts a limit on the expected value $N_{\rm dv}$ for which the resulting number of events is incompatible with the observed zero events for a given confidence level.
From that limit we derive the upper bound on the active-sterile neutrino mixing for benchmark masses between 2 and 30 GeV.
We choose for our analysis the 95\% C.L. which corresponds to the limit $N_{\rm dv}\geq 3.09$ events, as is discussed in ref.~\cite{Feldman:1997qc}.
The above discussed procedure can in principle be adjusted to other detectors, to other experiments at other colliders, and to a different statistical treatment.

\paragraph{Constraints:}
Since LHCb's analysis was conducted for long lived particles decaying to $\mu$ and jets, we consider only heavy neutrino decays into $\mu jj$ and we apply the same cuts to our signal sample: 
\begin{itemize}
\item $N(\mu)=1$ and $N(j)>0$
\item $2<\eta(f)<5$, $f = \mu,j$
\item $P_t(\mu)>12$ GeV
\item $M[\mu jj]> 4.5$ GeV
\end{itemize}
In the last cut, $M[\mu jj]$ corresponds to the reconstructed invariant mass of the heavy neutrino, and the cut rejects signal events from heavy neutrinos with masses below 4.5 GeV. 
We note that we have not simulated this reconstruction and just set this invariant mass equal to the heavy neutrino mass. 
Furthermore, we impose the cut $\vartheta[\mu jj]<0.34$ due to the geometric acceptance of the LHCb detector, in order to save simulation time. 
In the following we assume that after LHCb's track reconstruction we pass the analysis' preselection cuts with four recorded tracks or more.

We carry out the analysis for the special case $|\theta_e|=|\theta_\tau|= 0$ and $|\theta_\mu|\neq 0$, such that Br$_{\mu jj} \simeq 0.55$ (neglecting the small dependency on the mass $M$), noting that the $N$ also decays via $Z$ and $h$.  
We consider the displaced heavy neutrino decays to take place in two different regions:
\begin{itemize}
\item Region 1 restricts the decays to be inside the VELO, for which we take $r_{\rm max}=50$~cm, $z_{\rm max}=40$~cm and $r_{\rm min}=2$~cm. Here we assume a signal reconstruction efficiency of 100\%.
\item Region 2 restricts the decays to be inside the volume with $r_{\rm max}=60$~cm and $z_{\rm max}=2$~m, which is the radial extension and distance to the TT tracking station, and $r_{\rm min}=5$~mm. For radial displacements between 5~mm and 2~cm, due to, e.g., detector related backgrounds and blind spots, we use a 50\% signal reconstruction efficiency \cite{AurelioBay}. We also use 50\% efficiency for longitudinal displacements between 40~cm and 2~m\cite{AurelioBay}. Region 2 also includes region 1, where a signal reconstruction efficiency of 100\% is used.
\end{itemize} 

\noindent We remark that a realistic efficiency is dependent on the mass $M$ and that it decreases for masses towards 4.5~GeV. Therefore, a detailed analysis is necessary, which is beyond the scope of this paper.

For the calculation of the number of events $N_{\text{x}N}$ of the different processes $pp \to \text{x} N \to \text{x} (\mu j j)_{\rm displaced}$ with $\text{x}=\{\nu,\ell^\pm\}$ we use the symmetry protected model discussed above.
Note that for $\text{x}=\ell^\pm$ the event has a prompt charged lepton, which is of no consequence since the LHCb analysis has not put a veto on prompt charged leptons.
We evaluate the cross sections for all the three processes with the above applied selection cuts by using WHIZARD \cite{Kilian:2007gr,Moretti:2001zz} with the parton distribution function set to CTEQ6L. We note that we neglected theoretical uncertainties and uncertainties on the input parameters.
We show representatively the cross sections for heavy neutrino masses $M$ of 5 GeV in tab.~\ref{tab:cross sections} for the center-of-mass energies of 7 and 8 TeV, and also 13 TeV for comparison. 
We note that for heavy neutrino masses $M$ around 10 GeV, the production cross sections do not vary much with the mass.
\begin{table}
\begin{center}
\begin{tabular}{c| c c c}
$\sqrt{s}$ & $\frac{\sigma_{\nu N}\,{\rm Br}_{\mu jj}}{\theta^2}$ & $\frac{\sigma_{\ell^- N}\,{\rm Br}_{\mu jj}}{\theta^2}$ & $\frac{\sigma_{\ell^+ N}\,{\rm Br}_{\mu jj}}{\theta^2}$\\
\hline
7 TeV & 126 pb & 193 pb & 342 pb\\
8 TeV & 156 pb & 257 pb & 441 pb\\
13 TeV & 308 pb & 513 pb & 939 pb\\%
\end{tabular}
\end{center}
\caption{Base signal cross sections in the fiducial volume covered by LHCb. They are given by the Drell-Yan production cross sections of heavy neutrinos times the branching ratio into $\mu jj$ and divided by the square of the active-sterile neutrino mixing angle.
A benchmark heavy neutrino mass of 5 GeV and $|\theta_e|=|\theta_\tau|=0$ was adopted. 
For the evaluation, WHIZARD with the CTEQ6L parton distribution function have been used.}
\label{tab:cross sections}
\end{table}

\begin{figure}
\centering
\includegraphics[width=0.4\textwidth]{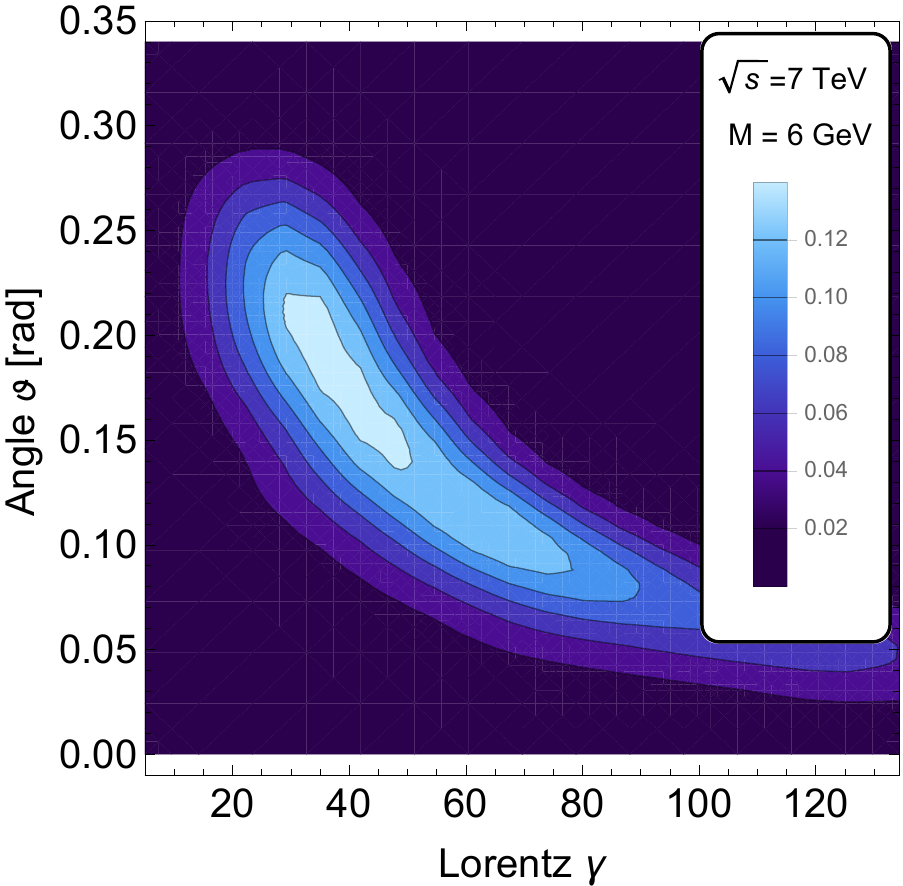}
\caption{An example for the used probability distributions $D_{\nu N}(\vartheta, \gamma)$, i.e. a produced heavy neutrino flying off with polar angle $\vartheta$ with corresponding Lorentz factor $\gamma$.}
\label{fig:pdfs}
\end{figure}

We construct the probability distribution $D_{\text{x}N}(\vartheta,\gamma)$ in eq.~\eqref{eq:masterequation} from simulated Monte Carlo event samples with $10^4$ events per benchmark mass for each of the three different processes, which were generated with WHIZARD at the parton level. An example distribution is shown in fig.\ \ref{fig:pdfs}.
Moreover, the decay width $\Gamma_N$ for different neutrino parameters is also obtained numerically with WHIZARD.

\begin{figure}

\begin{picture}(100,150)
\put(0,0){\includegraphics[width=0.4\textwidth]{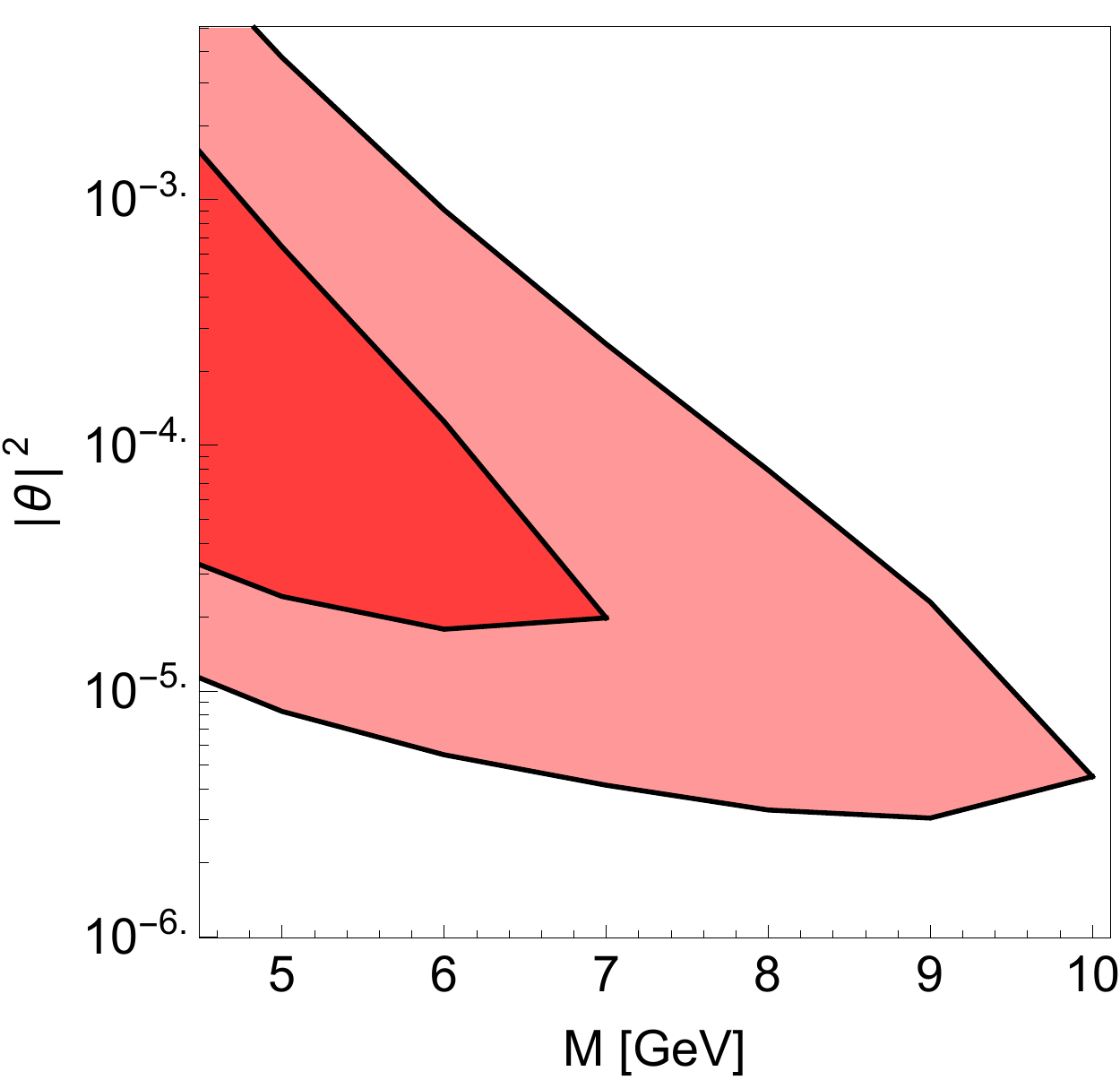}}
\put(100,145){\colorbox{white}{\includegraphics[width=0.175\textwidth]{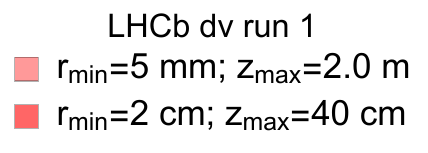}}}
\end{picture}
\caption{
Estimated exclusion limits at 95\% C.L. for the active-sterile neutrino mixing parameter $|\theta|^2$ as function of $M$ from the recent LHCb search for displaced vertices (dv) \cite{Aaij:2016xmb}. The dark red area is a conservative estimate considering only the region with radial displacement above 2 cm and longitudinal displacement below 40 cm, where no events have been observed at all and where no loss in signal reconstruction efficiency is expected. The light red area uses the region 2 of the detector volume, as described in the main text, with estimates for the efficiency in the additional regions. The limits are shown for $|\theta|^2 = |\theta_\mu|^2$ (i.e.\ $|\theta_e|=|\theta_\tau|= 0$).
}
\label{fig:sensitivity}
\end{figure}

We define the region of parameter space that is excluded by the LHCb  search for displaced vertices via the condition:
\begin{equation}
N_{\rm dv}(7 \text{ TeV},M,|\theta|^2)+N_{\rm dv}(8 \text{ TeV},M,|\theta|^2)\geq 3.09\,.
\end{equation}
As discussed at the end of the previous section this ensures that parameters fulfilling this condition yield at least one event that is incompatible with the observation of LHCb, and are thus excluded at the 95\% confidence level.
Our estimate for the LHCb exclusion limit for sterile neutrino parameters from displaced vertex searches is shown in fig.~\ref{fig:sensitivity}. The dark red region in the plot is a conservative estimate for the exclusion limit where only region 1 is used. For the limit shown in light red we use region 2, with estimated signal reconstruction efficiency of 50\% outside region 1.

In the case of $|\theta_e| \neq 0 \neq |\theta_\tau|$, the number of events $N_{\text{x}N}$ is rescaled with the branching ratio Br$_{\mu jj} = 0.55 |\theta_\mu|^2/|\theta|^2$, while the integral part of eq.~\eqref{eq:masterequation} is dependent on the mixing $|\theta|^2$.

\paragraph{Sensitivity forecasts:}
Finally, we project the sterile neutrino sensitivities of LHCb for run 2 at 13 TeV with 5 fb$^{-1}$, and for the high-luminosity phase with a total integrated luminosity of 380 fb$^{-1}$.
For the projected sensitivities the underlying assumption is that the background free environment remains the same as for run 1, and we also adopt the case with $|\theta|^2 = |\theta_\mu|^2$, i.e.\ $|\theta_e| = |\theta_\tau| = 0$.

\begin{figure}

\begin{picture}(100,150)
\centering
\put(0,0){\includegraphics[width=0.4\textwidth]{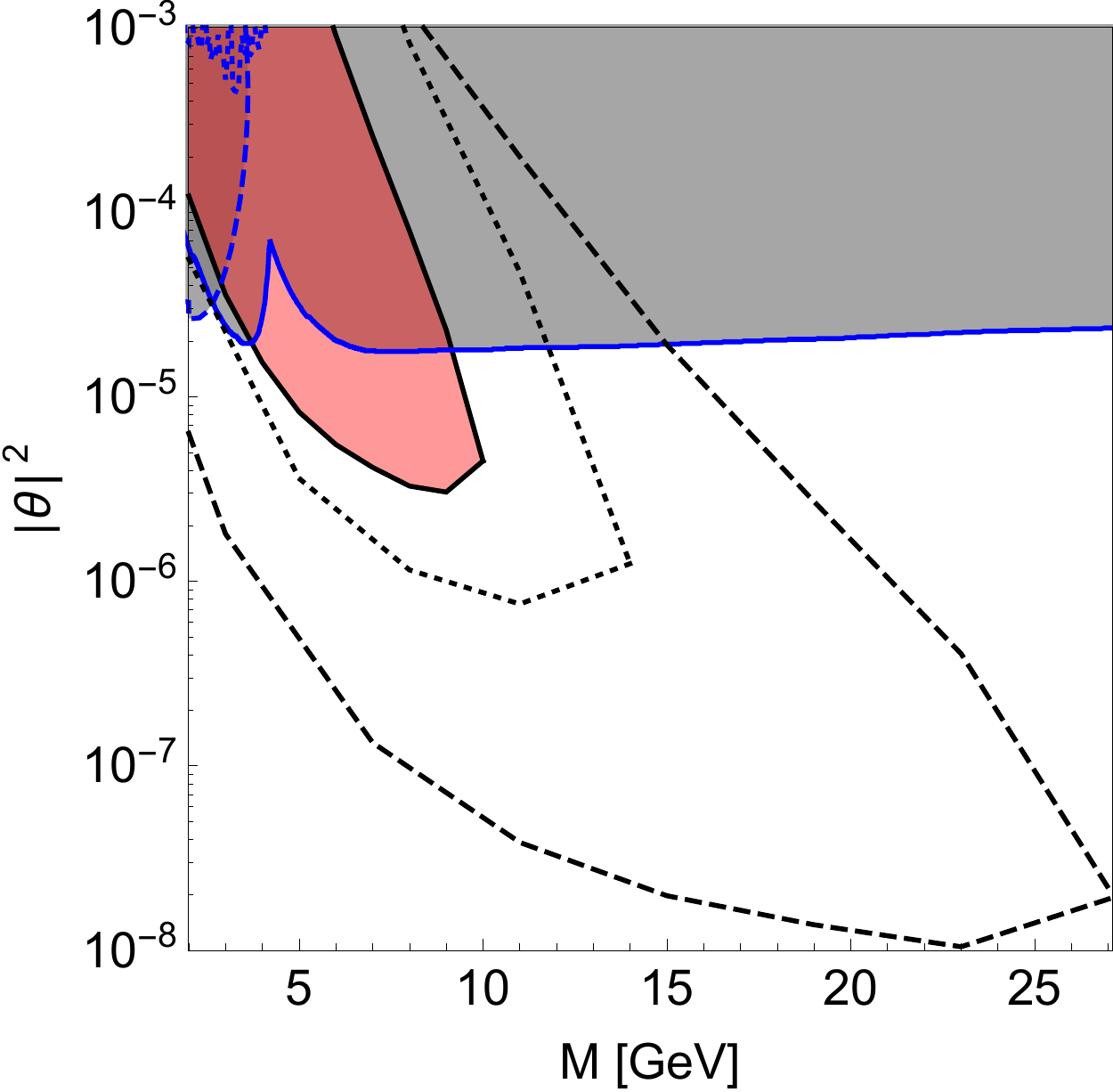}}
\put(139,107){\colorbox{white}{\includegraphics[width=0.1\textwidth]{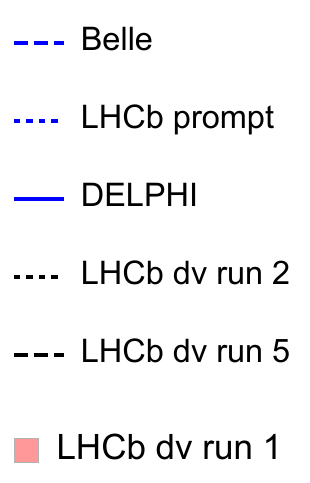}}}
\end{picture}

\caption{Projected 95\% C.L. sensitivities to the active-sterile neutrino mixing parameter $|\theta|^2$ as function of $M$ from LHCb searches for displaced vertices (dv) with muons plus jets at 13 TeV, compared to the present limits from DELPHI \cite{Abreu:1996pa}, Belle \cite{Liventsev:2013zz} and from the LHCb prompt \cite{Aaij:2014aba,Shuve:2016muy} and LHCb displaced vertex searches (cf.\ fig.\ \ref{fig:sensitivity}). The black dotted and dashed lines represent the sensitivities for the present amount of data of 5 fb$^{-1}$ (dv run 2) and the expected 380 fb$^{-1}$ (dv run 5) for the high-luminosity run. The sensitivities are shown for $|\theta|^2 = |\theta_\mu|^2$ (i.e.\ $|\theta_e|=|\theta_\tau|= 0$). 
Note that in the current LHCb displaced vertex analysis events with $M<4.5$ GeV would have been removed by cuts. We nevertheless show this region for comparison.}
\label{fig:sensitivity-forecast}
\end{figure}

We show the resulting projected 95\% C.L.\ sensitivities in fig.~\ref{fig:sensitivity-forecast}, in comparison to the estimated  present exclusion limit of fig.~\ref{fig:sensitivity} and other existing limits.\footnote{We note that the other analyses consider a simplified model with only one sterile neutrino, which strictly speaking yields a too large mass of the light neutrino. We can nevertheless compare our constraint derived in the SPSS model with these bounds, since for the considered processes the heavy neutrino production cross section and the kinematics are identical.} These include the 95\% C.L.\ limit from DELPHI \cite{Abreu:1996pa}, obtained from a combination of mono jet and acollinear jets searches for a short and long lived heavy neutral lepton, the limit from Belle \cite{Liventsev:2013zz}, derived from $B$-meson decays at the 90\% C.L., and the LHCb 95\% C.L.\ limit \cite{Aaij:2014aba}, also derived from $B$-meson decays (see ref.~\cite{Shuve:2016muy} for a revised limit).  
Furthermore, the active-sterile mixing parameters are also constrained indirectly from precision measurements, where, e.g.\ for $M \sim 10$ GeV, the constraint on $|\theta_\mu|^2$ is $\simeq10^{-4}$ \cite{Antusch:2015mia} (for similar discussions see also, e.g., \cite{Fernandez-Martinez:2016lgt,Das:2017nvm}).

We notice that, for 4.5 GeV $ < M <$ 10 GeV, our estimates suggest that already the present LHCb displaced vertex limit (cf.\ fig.\ \ref{fig:sensitivity}) provides the strongest bound on the parameter $|\theta|^2 = |\theta_\mu|^2$ (i.e.\ for $|\theta_e|=|\theta_\tau|= 0$). 
Furthermore, we find that future LHCb displaced vertex searches have the potential to significantly improve the sensitivity on $|\theta|^2$. 
Forecasts for the other LHC experiments can be found for instance in \cite{Izaguirre:2015pga} and feature comparable sensitivities.   
For the projected sensitivity for 2 GeV $ < M <$ 4.5 GeV, it will be interesting to explore whether the invariant mass cut on the tracks from the displaced vertex can be relaxed.

\paragraph{Conclusions:}
We have analysed the sensitivity of displaced vertex searches at LHCb for testing sterile neutrinos. Using the recently published displaced vertex search at LHCb based on run 1 data, we derived estimates for the constraints on sterile neutrino parameters. Our estimates indicate that for sterile neutrino masses $M$ around $9$ GeV the active-sterile neutrino mixing is constrained down to $\sim 3 \times 10^{-6}$ (at 95\% C.L.). They suggest that for 4.5 GeV $< M <$ 10 GeV, already the currently analysed  LHCb data provides the strongest present exclusion limit for active-sterile neutrino mixing. 

Furthermore, we presented forecasts for the sensitivities that could be obtained from the run 2 data and also for the high-luminosity phase of the LHC. These future searches could be able to probe the active-sterile mixing angles down to $\sim 10^{-8}$ (at 95\% C.L.) for sterile neutrino masses $M$ around $20$ GeV, more than three orders of magnitude better than the present bounds.  

We remark that searches for displaced vertices with different signatures, i.e.\ $N \to e jj$, $N \to \tau jj$, are desirable to complement existing analyses because they test the other active-sterile mixing parameters $|\theta_e|$ and $|\theta_\tau|$.
Moreover, we expect that a dedicated analysis for the process $pp \to \ell_\alpha N$ could reduce the $b$-quark backgrounds with radial displacements below 5 mm, leading to improved sensitivities. This could be achieved for instance by selecting, or even triggering, on the prompt lepton in addition to the displaced vertex. Thereby selecting prompt lepton momenta above 6 GeV (15 GeV) reduces the signal sample from charged Drell-Yan production by only 6\% (20\%). A dedicated analysis would also allow to infer a realistic signal reconstruction efficiency for displaced vertices in any of LHCb's detector components.  

In summary, the LHCb is an excellent environment for sterile neutrino searches via displaced vertices. 

\paragraph{Acknowledgements:} 
This work has been supported by the Swiss National Science Foundation. O.F.\ acknowledges support from the ``Fund for promoting young academic talent'' from the University of Basel under the internal reference number DPA2354. We thank Martino Borsato for stimulating discussions on the LHCb results. We are indebted to Aurelio Bay for his invaluable support with the LHCb experimental setup and careful reading of the manuscript.

\bibliographystyle{unsrt}

\end{document}